\documentclass[aps,showpacs,preprintnumbers,amsmath,amssymb�eqsecnum,
twocolumn, tightenlines,
]{revtex4}

\usepackage[dvips]{graphicx}
\usepackage{bm}
\usepackage{longtable}
\usepackage{amsmath}
\usepackage{dcolumn}
\usepackage{placeins}

\begin{document}

\bibliographystyle{apsrev}

\title{Transition amplitudes, polarizabilities and energy levels
  within optical wavelength of highly charged ions Sm$^{14+}$ and
  Sm$^{13+}$.} 

\author{A. Kozlov}\email[email:]{o.kozloff@student.unsw.edu.au}
\author{V. A. Dzuba}\email[email:]{dzuba@phys.unsw.edu.au}
\author{V. V. Flambaum}\email[email:]{flambaum@phys.unsw.edu.au}

\affiliation{School of Physics, University of New South Wales, Sydney
  2052, Australia}

    \date{\today}

    \begin{abstract}
We discuss possible search for optical transitions in Sm$^{13+}$ and Sm$^{14+}$ using
{\em ab initio} calculations of differential dynamic
polarizability. We calculate  dynamic polarizability for  M1 transition
between first and second excited states of  Sm$^{14+}$ . Transition amplitudes and
energies within optical range for  states that contribute to the 
polarizability of the mentioned transition are presented. Employing
simple analytical formula for polarizability data in the vicinity of a
resonance and assuming that several values of the polarizability for
different laser frequencies will be measured one can find the accurate
position of the resonance. Results of similar calculations of
amplitudes and energies of states that contribute to the polarizability of
the M1 transition between ground and first excited states of
Sm$^{13+}$ are also presented. 
    \end{abstract}


\maketitle

\section{Introduction}

The physics of highly charged ions (HCI) has long and rich history due
to the role the ions play in studying laboratory and cosmic plasma. 
Recently, the interest to the subject was further elevated due to
proposals to use HCI for exceptionally accurate atomic
clock~\cite{clock1,clock2} and for laboratory search for possible time
variation of the fine structure
constant~\cite{var1,var2,var3,var4}. The latter proposal suggests the
use of HCI with optical transitions between states of different
configurations. The existence of such transitions is due to level
crossing while moving from Madelung to Coulomb level ordering along an
isoelectronic sequence with increasing nuclear charge
$Z$~\cite{optHCI}. One of the main obstacles in the use of HCI with
optical transitions is absence of experimental data on the spectra of
the ions. Theoretical calculations are also difficult just because of
level crossing. Level crossing means that the energy interval between
states of different configurations is very small ($\sim 10^{-2} - 10^{-3}$) compared to the total
ionization energy of valence electrons. As a result, the relative theoretical error in this interval is enhanced  $\sim 10^{2} - 10^{3}$ times. For example, different calculations give
different ground states for Sm$^{14+}$, Eu$^{14+}$,
etc.~\cite{optHCI,Safronova}.  

Experimental study of the optical transitions is likely to have
problems as well. All these transitions are very weak magnetic dipole
(M1), electric quadrupole (E2) or strongly suppressed electric dipole
(E1) transitions. This is because level crossing in HCI happens mostly
between $s$ and $f$ levels or $p$ and $f$ levels. The $s-d$ crossing
happens for low ionization degree and there is no level crossing
consistent with selection rules for electric dipole
transitions. However, electric dipole optical transitions are still
possible between many-valence-electron states of HCI due to
configuration mixing. These transitions are suppressed because leading
configurations do not contribute to the amplitude and small admixture
of appropriate configurations make the electric dipole transition
possible. In this paper we suggest to employ dynamic Stark shift of
single known transition for recovering other optically accessible
transitions.  

We consider optically accessible transitions in Sm$^{14+}$ and
Sm$^{13+}$ ions. For these ions it is reasonable to consider M1
transitions between first and second exited states (first exited state
is metastable) for Sm$^{13+}$ and ground to first exited state
transition in Sm$^{14+}$. If differential dynamic Stark shift of such
transitions in external laser field is measured for different
frequencies of the light the information about ion spectra can be
extracted. The value of the
dynamic Stark shift is determent by E1 or M1 transitions from these two
states to other states. Therefore, studying the
dependence of the shift on frequency of the laser field can reveal the
positions of theses other states. It is important that in contrast to
direct scanning we don't have to come close to the resonance while its
position can be found with very high accuracy. This may have
significant advantage in searching for weak transitions in considered
HCI. 

\section{Calculations}
\subsection{Energy levels}
The results of calculations presented in this paper rely on method
described in details in \cite{Dzuba:1996, Dzuba:2005,Ginges:2006}. A
brief description of this method is presented below.   

We use the $V^{N-M}$ approximation~\cite{Dzuba:2005}. The core
electron states were obtained in Hartree-Fock approximation for $N-M$
electrons, where $N$ 
and $M$ are total number of electrons and number of electrons above
closed shells ("valence electrons"). The Hartree-Fock (HF) Hamiltonian of
the system has the form
\begin{equation}\label{HF}
\hat H_{HF}=\sum_{i=1}^{M}c\alpha \hat {\bf p_i} +
(\beta-1)mc^2-\frac{Ze^2}{r_i}+V^{N-M}(r_i), 
\end{equation}
where ${\bf \hat p_i}$ and ${\bf r_i}$ are operator of momentum and
coordinate of electron, $V^{N-M}$ is the self-consistent HF potential.

The configuration interaction method combined with the many-body
perturbation theory (the CI+MBPT method~\cite{CI+MBPT} is used to
construct the many-electron states for valence electrons. The
effective CI Hamiltonian has the form 
\begin{equation}\label{H_eff}
\hat H^{CI}=\sum_{i=1}^{M}\hat h_1(r_i)+\sum_{j>i=1}^{M}\hat h_2(r_i, r_j),
\end{equation}
where $\hat h_1(r)$ is the single-electron operator and $\hat h_2(r_i,
r_j)$ is the two-electron operator. The single electron operator $\hat
h_1(r)$ differs from (\ref{HF}) by an extra operator $\Sigma_1(r)$
\begin{equation}\label{h1}
\hat h_1(r_i)=c\alpha \hat {\bf p_i} +
(\beta-1)mc^2-\frac{Ze^2}{r_i}+V^{N-M}(r_i)+\Sigma_1(r_i). 
\end{equation}
This $\Sigma_1$ operator represents The correlation interaction between a
particular valence electron and electrons in the core.
The two electron part of (\ref{H_eff}) is given by 
\begin{equation}
\hat h_2(r_i, r_j)=\frac{e^2}{|{\bf r_i}-{\bf r_j}|}+\Sigma_2(r_i, r_j),
\end{equation}
where $\Sigma_2$ accounts for screening of Coulomb interaction between
valence electrons by core electrons. We calculate the $\Sigma_1$ and
$\Sigma_2$ operators in the lowest, second order of the MBPT.  

The CI many-electron wave function is written in a form
\begin{equation}\label{func}
\Psi=\sum_k c_k \Phi_k(r_1,...,r_M),
\end{equation}
where  $\Phi_k$ are determinants
made of single electron eigenfunctions of (\ref{HF}) combined in a way
to have appropriate value of total angular moment $J$. The expansion
coefficients $c_k$ and corresponding energies are found by solving the
matrix eigenvalue problem
\begin{equation} \label{CIM}
  H^{CI}\Psi = E\Psi
\end{equation}
for lowest states of definite $J$ and parity.

\subsection{Transition amplitudes and dynamic polarizabilities}

Electric dipole transition amplitudes are calculated using the
time-dependent Hartree-Fock method~\cite{Dzuba:1987} (equivalent to
the well-known random-phase approximation) and the CI technique 
\begin{equation}\label{E1}
\left<a|E1|b\right>=\left<\Psi^{(a)}|d_z+\delta V^{N-M}_{ij}|\Psi^{(b)}\right>,
\end{equation}
where $d_z=-ez$ is the $z$-component of the dipole moment operator and
$\delta V^{N-M}$ is the
correction to core potential due to its polarization by
external electric field. Electron wavefunctions $\Psi^{(a)}$ and
$\Psi^{(b)}$ were obtained using the CI+MBPT techniques described in
previous section. 

Dynamic Stark shift is considered in the Appendix. For our analysis we
need scalar polarizability given by (\ref{1Pol}). Its calculation
involves summation over complete set of intermediate many-electron
states. We use the Dalgarno-Lewis method~\cite{Dalgarno:1955} to
reduce this summation to solving a system of linear equations with the
CI matrix. The  polarizability is rewritten as 
\begin{equation}
\alpha(\omega)=\frac{2}{3(2J+1)}\sum_{J'=J,J\pm 1}\left<\delta\Psi_{J'}^{(a)}||{\bf
    d}||\Psi_J^{(a)}\right>, 
\end{equation}
where $J$ is the total angular momentum of the state $a$. 
The correction $\delta\Psi_{J'}^{(a)}$ to the wavefunction
$\Psi_{J}^{(a)}$ due to the laser electric field is found from the equation
\begin{equation}
\left(H^{CI}-E_a\right)\delta \Psi_{J'}^{(a)} = -\left(d_z+\delta
  V^{N-M}\right)\Psi_{J}^{(a)}. 
\end{equation}
The same formulas can be used to calculate energy shift in laser
magnetic field.

\FloatBarrier
\section{Energy levels of Sm$^{14+}$ and Sm$^{13+}$.}

\begin{table*}[t]\center
\caption{E1 and M1-allowed transitions within the optical wave-lengths from one of the reference levels of Sm$^{14+}$.}
{\renewcommand{\arraystretch}{0}%
\begin{tabular}{c l c l c c c}
\hline\hline
\rule{0pt}{4pt}\\
\multicolumn{2}{c}{initial state} & \multicolumn{2}{c}{final state} & transition & matrix & partial width, \\
\rule{0pt}{2pt}\\
$J_i$ & $E_i, $cm$^{-1}$ & $J_f$ & $E_f, $cm$^{-1}$ & energy, & element & $\Gamma$, a.u.\\
\rule{0pt}{2pt}\\
 & & & & a.u. & $\left<i||E1||f\right>, a_0$ &\\
\rule{0pt}{4pt}\\
\hline
\rule{0pt}{4pt}\\
2 & 1243 & 2 & 9337 & 0.0371 & $1.1\times10^{-2}$ & $6.3\times10^{-16}$\\
\rule{0pt}{2pt}\\
2 & 1243 & 3 & 13070 & 0.0539 & $-2.1\times10^{-3}$ & $7.2\times10^{-17}$\\
\rule{0pt}{2pt}\\
3 & 3053 & 2 & 9337 & 0.0288 & $-6.6\times10^{-3}$ & $7.6\times10^{-17}$\\
\rule{0pt}{2pt}\\
3 & 3053 & 3 & 13070 & 0.0456 & $-1.2\times10^{-2}$ & $1.0\times10^{-15}$\\
\rule{0pt}{2pt}\\
3 & 3053 & 4 & 0 & 0.0139 & $-0.95\times10^{-3}$ & $1.8\times10^{-19}$\\
\rule{0pt}{2pt}\\
3 & 3053 & 4 & 13891 & 0.0494 & $3.9\times10^{-3}$ & $1.0\times10^{-16}$\\
\rule{0pt}{2pt}\\
3 & 3053 & 4 & 21048 & 0.0820 & $1.4\times10^{-3}$ & $6.22\times10^{-17}$\\
\rule{0pt}{4pt}\\
\hline
\rule{0pt}{4pt}\\
\multicolumn{2}{c}{initial state} & \multicolumn{2}{c}{final state} & transition & matrix & partial width, \\
\rule{0pt}{2pt}\\
$J_i$ & $E_i, $cm$^{-1}$ & $J_f$ & $E_f, $cm$^{-1}$ & energy, & element & $\Gamma$, a.u.\\
\rule{0pt}{2pt}\\
 & & & & a.u. & $\left<i||M1||f\right>, a_0$ &\\
\hline
\rule{0pt}{4pt}\\
2 & 1243 & 3 & 14603 & 0.0609 & $-2.9\times10^{-3}$ & $9.4\times10^{-17}$\\
\rule{0pt}{2pt}\\
3 & 3053 & 3 & 14603 & 0.0526 & $-8.3\times10^{-4}$ & $5.0\times10^{-18}$\\
\rule{0pt}{2pt}\\
3 & 3053 & 4 & 8092 & 0.0230 & $9.0\times10^{-3}$ & $3.8\times10^{-17}$\\
\rule{0pt}{2pt}\\
\hline\hline
\end{tabular}}\label{Tab:3}
\end{table*}

\begin{table}\center
\caption{Energy spectrum of Sm$^{+14}$. The results were obtained using configuration interaction method and include up to second order of many body perturbation theory. Reference transition is allocated with bold font, measurement of its dynamic Stark shift gives necessary a differential polarizability can be obtained. Star symbol indicates levels that contribute to differential polarizability of reference transition and therefore can be calculated using proposed method.}
{\renewcommand{\arraystretch}{0}%
\begin{tabular}{l c c c c}
\hline\hline
\rule{0pt}{4pt}\\
Configuration & J & Parity & $\Delta E,\text{cm}^{-1}$ & g-factor\\
\rule{0pt}{4pt}\\
\hline
\rule{0pt}{4pt}\\
$4f^2$ * & 4 & e & 0 & 0.8\\
\rule{0pt}{2pt}\\
$\textbf{5s4f}$ & \textbf{2} & \textbf{o} & \textbf{1243} & \textbf{0.67}\\
\rule{0pt}{2pt}\\
$\textbf{5s4f}$ & \textbf{3} & \textbf{o} & \textbf{3053} & \textbf{1.07}\\
\rule{0pt}{2pt}\\
$4f^2$ & 5 & e & 5409 & 1.0\\
\rule{0pt}{2pt}\\
$5s4f$ * & 4 & o & 8092 & 1.25\\
\rule{0pt}{2pt}\\
$4f^2$ * & 2 & e & 9377 & 0.67\\
\rule{0pt}{2pt}\\
$4f^2$ & 6 & e & 10877 & 1.16\\
\rule{0pt}{2pt}\\
$4f^2$ * & 3 & e & 13070 & 1.0\\
\rule{0pt}{2pt}\\
$4f^2$ * & 4 & e & 13891 & 1.14\\
\rule{0pt}{2pt}\\
$5s4f$ * & 3 & o & 14611 & 1.0\\
\rule{0pt}{2pt}\\
$4f^2$ * & 4 & e & 21048 & 1.1\\
\rule{0pt}{2pt}\\
$5s^2$ & 0 & e & 30908 & 0.0\\
\rule{0pt}{4pt}\\
\hline\hline
\end{tabular}}\label{Tab:2}
\end{table}

All optical E1-transitions in highly charged ions are narrow. This is
because these transitions are in optical range due to $s-f$ or $p-f$
level crossing~\cite{optHCI}. The $f$ states are not connected to
either $s$ or $p$ states by electric dipole operator. However, if the
number of valence electrons is larger than one, the electric dipole
transition might be possible due to the mixing with appropriate
configuration. This mixing is small due to large energy intervals
between the states in HCI. For example, the $f^2$ -- $sf$ electric
dipole transition might be possible if second state is mixed with the
$df$ configuration. This mixing is small because it is inversely
proportional to the $d-f$ energy interval which is large in HCI. 

Let's consider in detail the search for electrical dipole optical
transitions in Sm$^{14+}$ ion. It has two valence electrons above
Xe-like core.  
There is a $4f-5s$ levels crossing for this ion~\cite{optHCI} which
means that all lower states of the ion are dominated by the $4f^2$,
$4f5s$ and $5s^2$ configurations and intervals between them are in
optical range. It makes this ion a candidate for optical clocks and
for experimental search of time variation of the fine structure
constant. 
Experimental spectrum of this ion is not known and we use {\em ab
  initio} calculations to find all data we need. The results for
energy levels and $g$-factors are presented in Table~\ref{Tab:2}. Note
that due to the level crossing energy intervals between states of
Sm$^{14+}$ are small compared to total two-electron removal
energy. Therefore, they are very sensitive to accurate treatment of
correlation and relativistic effects. For example, estimations of
Ref.~\cite{optHCI} give different order of states than those presented
in Table~\ref{Tab:2}. The most accurate calculations for Sm$^{14+}$
will be published elsewhere~\cite{Safronova}. Preliminary results of
~\cite{Safronova} indicate the same order of states as in present
work. 

\FloatBarrier
We consider differential dynamic polarizability in the M1 transition
between first and second excited states of Sm$^{14+}$. These states
are shown in Table~\ref{Tab:2} in bold. Both states are very
long-living states. 
Although, there is an allowed E1 transition from second exited state
to the ground state we expect it to be very weak for reasons discussed
above (see also calculated E1-transition amplitudes in
Table~\ref{Tab:3}). The first excited state can decay to the ground
state only via E3 transition. Due to its high order and small
frequency the probability of the transition is extremely small. 

\begin{figure*}
\includegraphics[width=0.85\textwidth]{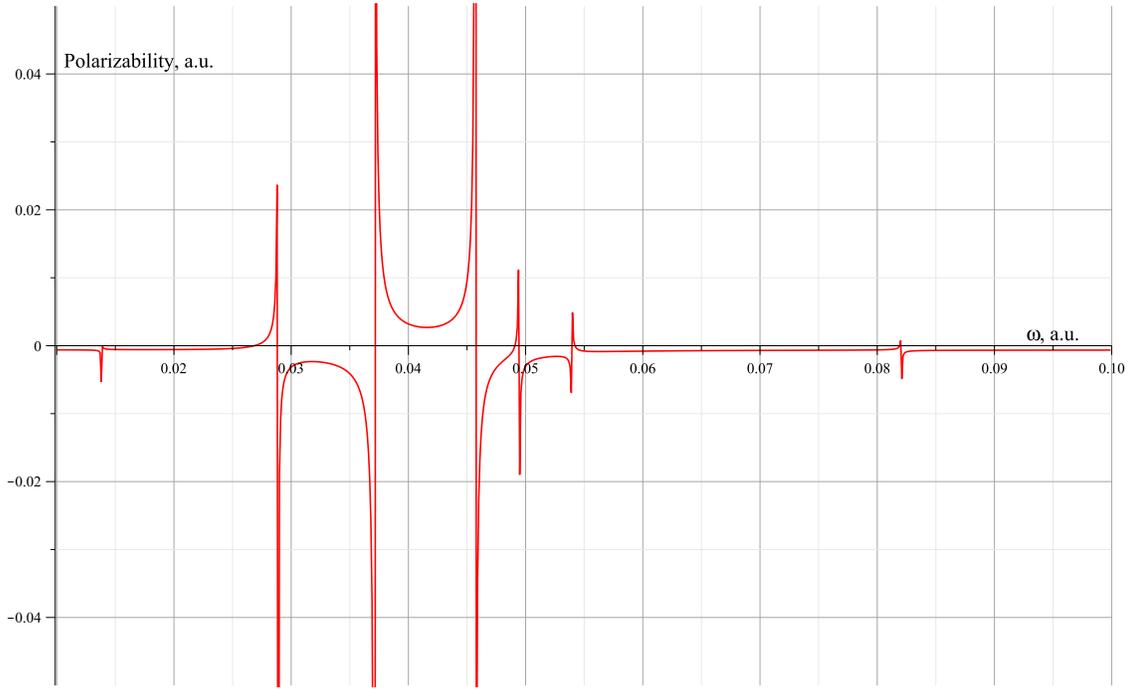}
\caption{Difference of the scalar polarizabilities  (\ref{1Pol}) for  Sm$^{+14}$ M1 transition. The resonances appear for $\omega$ equal to 0.0139; 0.0288; 0.0371; 0.0456; 0.0494; 0.0539; 0.0820.}\label{Fig:3}
\end{figure*}

Figure \ref{Fig:3} presents results of calculation of differential
polarizability of M1 transition between first and second exited states
in Sm$^{+14}$ (reference transition). Energy levels within the optical
range which contribute to the polarizability of the reference
transition are listed in Table~\ref{Tab:3}.   

\FloatBarrier
\begin{table*}\center
\caption{Energy levels of Sm$^{14+}$ recovered from data on
  Fig. \ref{Fig:4}. $\Delta E$ and $A$ are the interpolation
  parameters in eq. (\ref{Appr}). Sign of $A$ together with
  theoretical calculation results allows to pick correct one (bold) of
  two possible values of $E_k$ - exited energy level, that contributes
  to reference transition differential polarizability.} 
{\renewcommand{\arraystretch}{0}%
\begin{tabular}{c c c l}
\hline\hline
\rule{0pt}{4pt}\\
$\Delta E$, a.u. & $A/2(3J+1),  a_0^2$ & $\left<i||E1||f\right>^2, a_0^2$ & $E_k$, cm$^{-1}$\\
\rule{0pt}{4pt}\\
\hline
\rule{0pt}{4pt}\\
0.0139 & $+9.07\times10^{-7}$ & $9.02\times10^{-7}$ & \textbf{3053-3052=1}\\
\rule{0pt}{2pt}\\
& & & 1243+3052=4295\\
\rule{0pt}{2pt}\\
0.0288 & $-4.34\times10^{-5}$ & $4.36\times10^{-5}$ & \textbf{3053+6324=9377}\\
\rule{0pt}{2pt}\\
& & & 1243-6324=-5081\\
\rule{0pt}{2pt}\\
0.0371 & $+1.22\times10^{-4}$ & $1.21\times10^{-4}$ & \textbf{1243+8146=9389}\\
\rule{0pt}{2pt}\\
& & & 3053-8146=-5093\\
\rule{0pt}{2pt}\\
0.0456 & $-1.39\times10^{-4}$ & $1.44\times10^{-4}$ & \textbf{3053+10013=13066}\\
\rule{0pt}{2pt}\\
& & & 1243-10013=-8830\\
\rule{0pt}{2pt}\\
0.0494 & $-1.52\times10^{-5}$ & $1.50\times10^{-5}$ & \textbf{3053+10847=13900}\\
\rule{0pt}{2pt}\\
& & & 1243-10847=-9604\\
\rule{0pt}{2pt}\\
0.0539 & $+4.43\times10^{-6}$ & $4.41\times10^{-6}$ & \textbf{1243+11835=13078}\\
\rule{0pt}{2pt}\\
& & & 3053-11835=-8782\\
\rule{0pt}{2pt}\\
0.0820 & $-2.03\times10^{-6}$ & $1.96\times10^{-6}$ & \textbf{3053+18005=21058}\\
\rule{0pt}{2pt}\\
& & & 1243-18005=-16762\\
\rule{0pt}{4pt}\\
\hline\hline
\end{tabular}}\label{Tab:4}
\end{table*}

\begin{figure*}
\includegraphics[width=0.85\textwidth]{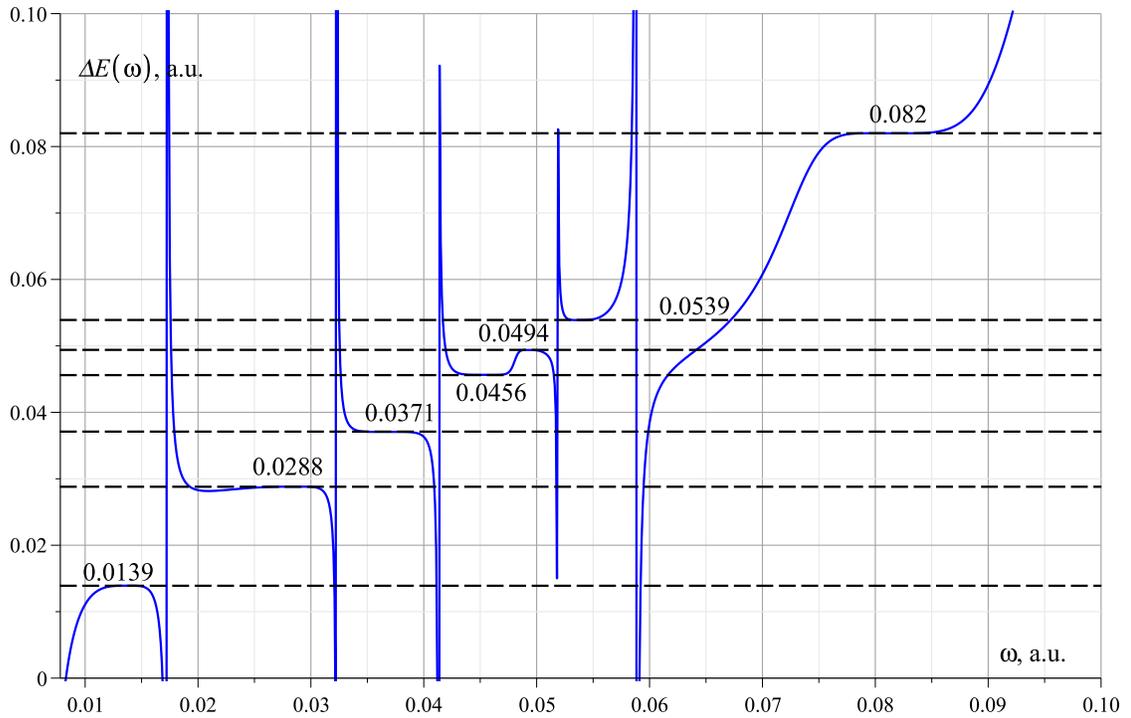}
\caption{Energy level position $\Delta E$ given by (\ref{DelE}) relative to one of the reference transition levels, as a function of external laser frequency $\omega_1$. Dashed lines corresponds to resonances in polarizability presented in Fig. \ref{Fig:3}. For small values of $\Delta \omega$ there is no sensitivity to $\Delta \omega$.}\label{Fig:4}
\end{figure*}

The results for relative position of the levels given by equation
(\ref{DelE}) are presented on Fig.~\ref{Fig:4}. Presence of horizontal
regions (same value of $\Delta E$ for different values of $\omega$) 
indicates the existence of frequency intervals where one resonance
strongly dominates. Fitting of the dynamic polarizability using
(\ref{DelE}) in these frequency intervals recovers the positions of
the energy levels which are in good agreement with direct
calculations. 
This means that such approximation for differential polarizability is
valid near resonances. The frequency intervals in which level can be
detected now are of the order of $10^{-4}$ a.u. 

There is an additional uncertainty in fitting procedure which needs to
be discussed. When differential polarizability is considered and
energy distance to the resonance $\Delta E$ is found from the fitting
procedure, it is not known to energy of which of two states this
$\Delta E$ should be added to find the position of the resonance
level. There is also a question about the sign of $\Delta E$. The sign
is always positive for the ground state polarizability.
For differential polarizability of excited states the sign of $\Delta
E$ must be consistent with the sign of $A$ (see (\ref{DelE})). This is
evident from comparing (\ref{pol}) and (\ref{Appr}). 
If $A<0$ then the energy of the resonance state is either $E_e +
\Delta E$ or $E_g - \Delta E$. If $A>0$ then the energy is $E_e -
\Delta E$ or $E_g + \Delta E$. The actual choice between these two
possibilities is easy when calculated spectrum is available. Note that
the accuracy of the calculations does not have to be very high since we
only need to choose between two very distinct possibilities.  

Table~\ref{Tab:4} illustrates reconstruction of the energy levels of
Sm$^{14+}$ from the data on the dynamic scalar polarizability of the
M1 transition (Fig.~\ref{Fig:3}). First two columns presents the
values of $\Delta E$ and $A$ obtained from (\ref{DelE}) using the
values of the polarizabilities close to corresponding resonance. The
last column of the table shows the recovering of the energies of the
resonance states using presumably known energies of the states for
which polarizability is measured and $\Delta E$ from first column. The
right choice of the sign of $\Delta E$ and to the energy of which of
the two states it should be added is shown in bold. The resulting
energies agree well with calculated energies of
Table~\ref{Tab:2}. Note that the energies of the states which
contribute to polarizabilities of both considered states are found
twice.  

Third column in Table~\ref{Tab:4} presents the squared reduced matrix
element of the electric dipole transition which can be compared with
the parameter $A$. In a single-resonance approximation they are
related by $|A| = \langle g ||E1||e \rangle^2/2(3J+1)$. One can see
from the table that they are really close in value. Some small
difference illustrates the accuracy of fitting by (\ref{Appr}). The
data in Table~\ref{Tab:4} shows that in the frequency intervals where
the fitting formula (\ref{Appr}) works well it can be used to recover
not only the energy positions of the resonance states but also the
values of electric dipole transition amplitudes. 

The procedure considered above implies dynamic Stark shift of
reference transition energy in external electric field of a laser. 
This shift is suppressed due to small values of electric dipole
transition amplitudes. The amplitudes are small because the
transitions cannot go between leading configurations and appear only
due to configuration mixing. On the other hand, there are magnetic
dipole transitions which are not suppressed  because they go
between states of the same configuration. In this situation magnetic
dipole transitions can give significant contribution to the dynamic
polarizability. To check this we have performed calculations of the M1
amplitudes for transitions which may affect the energy shift of the
reference transition. The results are presented in lower lines of
Table~\ref{Tab:3}.
To present M1 amplitudes we use the relation  Bohr magneton $\mu_B = \alpha/2 \approx 3.65 \times 10^{-3}$ a.u.
As one can notice the values of M1 and E1 amplitudes
are of the same order of magnitude. Therefore, they should be included
in the calculation of the total energy shift. 
The shift is described by the same equations as
ones presented in Appendix after replacing electric field with
magnetic in (\ref{Stark}) and E1 with M1 amplitudes in
(\ref{1Pol}). The analysis based on formula (\ref{Appr}) is still the
same. There are going to be
extra peaks on the graph of the energy shift as a function
of external frequency. This complicates the analysis, however the
positive side of this is that it allows to see more levels.
Theoretical calculations might be used to help identify the states
where the resonances originate from.
 
The same relation between optical E1 and M1 transitions is expected to be
valid for many HCI with more than one valence electron. In such
systems electric dipole transition amplitudes are small because they
cannot go between leading configurations and appear only due to
configuration mixing. On the other hand, there are always states of
the same configuration where M1 amplitudes are of the order of Borh
magneton. 

Table~\ref{Tab:5} presents the results of similar calculations for the 
Sm$^{13+}$ ion. This ion has one extra electron above closed shells
which leads to much larger number of transitions within the optical range.
The reference transition is the M1 transition between the ground
and first exited states with the energy of 6787 cm$^{-1}$.  
The last column of the table represents the
amplitudes, that can be used to reduce the number of fitting
parameters.  

For this ion there are only two levels of odd parity (reference
transition) within optical range. Therefore for the Sm$^{13+}$ ion there
will be no extra resonances in energy shift due to laser magnetic field 
as it was for the Sm$^{14+}$ ion. 
 
\begin{table}\center
\caption{E1-allowed transitions within the optical wave-lengths from one of the reference levels of Sm$^{13+}$.}
{\renewcommand{\arraystretch}{0}%
\begin{tabular}{c l c l c c}
\hline\hline
\rule{0pt}{4pt}\\
\multicolumn{2}{c}{initial state} & \multicolumn{2}{c}{final state} & transition & matrix \\
\rule{0pt}{2pt}\\
\multicolumn{2}{c}{$5s^2 4f^1$, odd} & \multicolumn{2}{c}{$5s^1 4f^2$, even} & energy, & element\\
\rule{0pt}{2pt}\\
 $J_i$ & $E_i, $cm$^{-1}$ & $J_f$ & $E_f, $cm$^{-1}$ & a.u. & $\left<i||E1||f\right>, a_0$\\
\rule{0pt}{4pt}\\
\hline
\rule{0pt}{4pt}\\
2.5 & 0 & 1.5 & 31974 & 0.1457 & $-3.5\times10^{-5}$\\
\rule{0pt}{2pt}\\
2.5 & 0 & 1.5 & 59831 & 0.2726 & $-7.9\times10^{-3}$\\
\rule{0pt}{2pt}\\
2.5 & 0 & 1.5 & 63794 & 0.2907 & $-3.4\times10^{-3}$\\
\rule{0pt}{2pt}\\
2.5 & 0 & 2.5 & 33648 & 0.1533 & $-3.2\times10^{-3}$\\
\rule{0pt}{2pt}\\
2.5 & 0 & 2.5 & 47679 & 0.2172 & $-1.4\times10^{-2}$\\
\rule{0pt}{2pt}\\
2.5 & 0 & 2.5 & 59004 & 0.2688 & $-5.9\times10^{-4}$\\
\rule{0pt}{2pt}\\
2.5 & 0 & 3.5 & 22824 & 0.1040 & $-2.1\times10^{-3}$\\
\rule{0pt}{2pt}\\
2.5 & 0 & 3.5 & 35940 & 0.1638 & $3.0\times10^{-3}$\\
\rule{0pt}{2pt}\\
2.5 & 0 & 3.5 & 44036 & 0.2006 & $1.1\times10^{-2}$\\
\rule{0pt}{2pt}\\
2.5 & 0 & 3.5 & 53901 & 0.2456 & $2.3\times10^{-3}$\\
\rule{0pt}{2pt}\\
3.5 & 6787 & 2.5 & 33648 & 0.1224 & $0.3\times10^{-3}$\\
\rule{0pt}{2pt}\\
3.5 & 6787 & 2.5 & 47679 & 0.1863 & $-2.5\times10^{-3}$\\
\rule{0pt}{2pt}\\
3.5 & 6787 & 2.5 & 59004 & 0.2379 & $6.8\times10^{-3}$\\
\rule{0pt}{2pt}\\
3.5 & 6787 & 3.5 & 22824 & 0.0731 & $0.3\times10^{-3}$\\
\rule{0pt}{2pt}\\
3.5 & 6787 & 3.5 & 35940 & 0.1328 & $2.0\times10^{-3}$\\
\rule{0pt}{2pt}\\
3.5 & 6787 & 3.5 & 44036 & 0.1697 & $-6.9\times10^{-3}$\\
\rule{0pt}{2pt}\\
3.5 & 6787 & 3.5 & 53901 & 0.2147 & $1.4\times10^{-2}$\\
\rule{0pt}{2pt}\\
3.5 & 6787 & 4.5 & 25357 & 0.0846 & $-3.0\times10^{-3}$\\
\rule{0pt}{2pt}\\
3.5 & 6787 & 4.5 & 37041 & 0.1378 & $1.2\times10^{-2}$\\
\rule{0pt}{2pt}\\
3.5 & 6787 & 4.5 & 39687 & 0.1499 & $3.0\times10^{-4}$\\
\rule{0pt}{2pt}\\
3.5 & 6787 & 4.5 & 46921 & 0.1829 & $1.7\times10^{-2}$ \\
\rule{0pt}{4pt}\\
\hline\hline
\end{tabular}}\label{Tab:5}
\end{table}

\FloatBarrier
\section{Conclusions}

It has been shown that the analysis of the
dynamic Stark shift for a single transition in HCI can be used to
recover a significant part of the spectrum of this ion as well as the
values of the electric dipole transition amplitudes between the
shifted states and states which contribute to their
polarizabilities. Highly charged ions Sm$^{14+}$ and Sm$^{13+}$
considered in the paper are of particular interest
since they are candidates for atomic clocks and for the search for time
variation of the fine structure constant. The ions have relatively
simple electron structure with two and three valence electrons above
closed shells. This makes it easier to base the analysis on the
theoretical calculations of the polarizabilities. However, similar
analysis based on experimental data is not limited to ions with simple
electron structure and can be useful for experimental study of wide
range of the HCI. 

\acknowledgments

The work was supported in part by the Australian Research Council.

\appendix
\section{Stark shift near resonance}

Energy shift of atomic levels in the presence of external electric field $\mathcal{E}$ of linearly polarized light with frequency $\omega$ can be written as~\cite{Manakov:1986,Manakov:1978}

\begin{equation}\label{Stark}
\Delta \varepsilon_{nJM}=-\left[\alpha_{nJ}^S(\omega)+\alpha_{nJ}^T(\omega)\frac{3M^2-J(J+1)}{J(2J-1)}\right]\frac{\mathcal{E}^2}{4},
\end{equation}
where $n, J, M$ are main quantum number, total electron angular momentum and its projection respectively and $\alpha_{nf}^S(\omega)$ and $\alpha_{nf}^T(\omega)$ are scalar and tensor dynamic polarizabilities of the state $n,J$. Averaging over all total angular momentum projections cancels out tensor polarizability, therefore for simplicity we will consider only scalar polarizability. It can be written as

\begin{equation}\label{1Pol}
\alpha_{nJ}^S(\omega)=\frac{2}{3(2J+1)}\sum_{n', J'=J-1}^{J'=J+1}\frac{\Delta E \langle nJ||\textbf{d}||n'J'\rangle^2}{\Delta E^2-\omega^2},
\end{equation}
where $\Delta E=E_n-E_{n'}$, $\textbf{d} = -e\textbf{r}$ is the electric dipole operator and summation goes over complete set of intermediate states. The above equation has singular points at $\omega=E_n-E_{n'}$, which correspond to the resonances. If frequency $\omega$ of the laser light is close to a resonance it is convenient to rewrite \ref{1Pol} in the following form

\begin{equation} 
\begin{split}
&\alpha_{nJ}^S(\omega)=-\frac{2}{3(2J+1)}\left(\frac{1}{2}\frac{\langle nJ||\textbf{d}||kJ'\rangle^2}{(E_n-E_k)-\omega}+\right.\\
&\frac{1}{2}\frac{\langle nJ||\textbf{d}||kJ'\rangle^2}{(E_n-E_k)+\omega}+\left.\sum_{n'\ne k} \frac{(E_n-E_{n'})\langle nJ||\textbf{d}||n'J'\rangle^2}{(E_n-E_{n'})^2-\omega^2}\right).
\end{split}\label{pol}
\end{equation}

Since $\omega$ is close to resonance energy $\Delta E=|E_n-E_k|$ first or second term in brackets determines behavior of $\alpha_0(\omega)$ depending on the sign of $\Delta E$. Hence for differential polarizability $\alpha^S(\omega)=\alpha_{n_1J_1}^S(\omega)-\alpha_{n_2J_2}^S(\omega)$ of the reference transition near resonance a simple analytical formula containing single resonance term and some simple approximation for the rest of the sum can be employed:

\begin{equation}\label{Appr} 
\Delta \alpha_{nJ}^S(\omega)=-\left(\frac{A}{\Delta E -\omega}+K\omega+C\right).
\end{equation}
Here $nJ$ is the one of the two states $n_1J_1$ or $n_2J_2$ which satisfy the resonance condition  $\omega \approx |E_{n'J'}-E_{nJ}|$; $A$, $K$, $C$ and $\Delta E$ are fitting parameters. It is assumed that $\Delta E>0$. Comparing (\ref{Appr}) to (\ref{1Pol}) one can see that the parameter $A$ is related to the electric dipole transition amplitude between the resonance states $nJ$ and $n'J'$ by $A=\pm \langle nJ||\textbf{d}||n'J'\rangle^2/3(2J+1)$. The plus sign corresponds to the case when $E_{n'J'} > E_{nJ}$, the minus sign is when
$E_{n'J'} < E_{nJ}$. Fitting measured differential Stark shift of the frequency of the reference transition as a function of the laser frequency using (\ref{Appr}) allows one to find the position of the resonance ($\Delta E$) and the value of the electric dipole transition amplitude between the states involved in the resonance ($A$). Note that there is still uncertainty due to the fact that it is still not known which of the the two reference states $n_1J_1$ or $n_2J_2$ is involved in the resonance. Fitting by (\ref{Appr}) does not distinguish between the two possibilities. One has to compare with the calculations or use some other considerations. For example, if $A<0$ then the state $nJ$ cannot be the ground state. More generally, it cannot be the state from which there is no electric dipole transitions to the lower states. 

It can be useful to have the formulae for the parameters $\Delta E$, $A$, $K$ and $C$ in (\ref{Appr}) for the case when the differential polarizability is known at four values of laser frequency, $\omega_1, \omega_2, \omega_3$ and $\omega_4$ separated by equal frequency intervals $\Delta \omega$. The formulae are 
\begin{eqnarray}\label{DelE}
\Delta E&=&\frac{\omega_4-Q\omega_1}{1-Q},\\
Q&=&\frac{\alpha_{nJ}^S(\omega_1) - 2\alpha_{nJ}^S(\omega_2) + \alpha_{nJ}^S(\omega_3)}{\alpha_{nJ}^S(\omega_2) - 2\alpha_{nJ}^S(\omega_3) + \alpha_{nJ}^S(\omega_4)} \nonumber \\
A&=&-\frac{\alpha_{nJ}^S(\omega_1) - 2\alpha_{nJ}^S(\omega_2) + \alpha_{nJ}^S(\omega_3)}{\Delta \omega^2}\frac{3(2J+1)}{4}\times \nonumber \\
&&(\Delta E - \omega_1)(\Delta E - \omega_2)(\Delta E - \omega_3).
\nonumber
\end{eqnarray}

\end{document}